\documentclass[twocolumn,showpacs,preprintnumbers,amsmath,amssymb]{revtex4}
\usepackage{graphicx}
\usepackage{dcolumn}
\usepackage{bm}

\def\deriv#1{\frac{\partial}{\partial#1}}
\newcommand{\js}{\mathbf{\hat j}^{\!\!~a}}
\newcommand{\sop}{s^{\!\!~a}}
\newcommand{\sd}{\vec S}

\newcommand{\ie}{{\it i.e.~}}
\newcommand{\eg}{{\it e.g.~}}
\newcommand{\half}{\frac{1}{2}}

\begin{document}

\title{Magnification of spin Hall effect in bilayer electron gas}

\author{Pei-Qing Jin and You-Quan Li}
\address{Zhejiang Institute of Modern Physics and Department
of Physics, Zhejiang University, Hangzhou 310027, P. R. China}

\begin{abstract}
Spin transport properties of a coupled bilayer electron gas with
Rashba spin-orbit coupling are studied.
The definition of the spin currents in each layer as well as
the corresponding continuity-like equations in the bilayer system are given.
The curves of the spin Hall conductivities obtained
in each layer exhibit sharp cusps around a particular
value of the tunnelling strength and the conductivities undergo
sign changes across this point.
Our investigation on the impurity
effect manifests that an arbitrarily small concentration of
nonmagnetic impurities does not suppress the spin Hall conductivity
to zero in the bilayer system.
Based on these features, an experimental scheme is suggested
to detect a magnification of the spin Hall effect.
\end{abstract}


\received{6 May 2007} \maketitle

\section{Introduction}\label{intro}

Manipulating the spin degree of freedom for electrons has recently
brought in an emerging information technology,
spintronics~\cite{Daykonov, Wolf, Zutic}, which offers novel clues
for designing devices based on traditional materials with
spin-related effects. In this promising field, the spin Hall
effect~\cite{Hirsch,Zhang,Niu0403} is regarded as a candidate method
to inject spin current in semiconductors. Based on the spin-orbit
coupling (SOC), an external electric field is required to drive a transverse
spin current while the magnetic field is not necessary, which is much
different from the traditional applications of the spin degree of
freedom.
A universal spin Hall conductivity $e/8\pi$ is predicted theoretically
in a clean single layer electron system~\cite{Niu0403}. Several groups'
calculations~\cite{Inoue,Halperin,Dimitrova,Raimondi,Rashba} showed that
nonmagnetic impurities would suppress this spin Hall conductivity to zero
while others indicated that the spin Hall conductivity is not zero
in the presence of  magnetic impurities~\cite{Inoue06,Wang}.
Experimentally, the spin accumulation in nonmagnetic
semiconductors has been observed~\cite{Wunderlich,Awschalom} and the
spin current was detected either by Kerr rotation
microscopy~\cite{Sih} or by two-color optical coherence control
techniques~\cite{Zhao}. Very recently, a direct electronic
measurement of the spin Hall effect has been reported~\cite{Valenzuela}
where the spin current induces the charge imbalance and a voltage is
detected.

As the SOC, which is crucial to the spin Hall effect, is a
relativistic effect and thus comparably weak, a natural question
is how to strengthen this effect. In the light of single layer
systems being considered in current literature, one may ask
whether a multi-layer system possesses a magnification effect and
what new phenomena will take place if the tunnelling between
layers is taken into account.
Another more realistic question is what will happen if there exist
impurities in a multi-layer system.

In this paper, we investigate the spin transport properties in a
coupled bilayer electron system with different SOC strengthes in
each layer as well as the tunnelling between layers. As a starting
point, we generalize the definitions of spin currents to a coupled
bilayer system and obtain the corresponding ``continuity-like''
equations. Carrying out calculations of the spin current
in the Heisenberg representation, we find that the spin Hall conductivity
in each layer manifests abrupt enhancement around a particular value
of the tunnelling strength between layers and undergoes a sign change
across this point. The influence of impurities is also studied. We indicate
that the spin Hall conductivity in the bilayer system can not be suppressed
to zero by an arbitrarily small concentration of impurities.
An experimental scheme is designed on the basis of these
features to magnify the spin Hall effect near the turning point.
Besides, possible logical gates are expected to be elaborated based on the sign
change of the spin current across this point.

The whole paper is organized as follows.
In Sec.~\ref{sec:continuity-like}, we generalize the
definition of the spin current in each layer and obtain the
``continuity-like'' equations.
In Sec.~\ref{sec:calculations},
the spin current as well as the spin Hall conductivity in each layer
are calculated in Heisenberg representation.
In Sec.~\ref{sec:disorder}, the influence of disorderly distributed
nonmagnetic impurities on the spin Hall conductivity is investigated.
In Sec.~\ref{sec:magnification}, we show the greatly enhanced
spin currents near the turning point and scheme out possible
experiment to detect a magnification of the spin Hall effect.
Finally, a brief summary is given in Sec.~\ref{sec:summary} and
some concrete expressions are written out in the appendix.

\section{``continuity-like'' equations}\label{sec:continuity-like}

As a proposition to study the spin transport, we firstly introduce
the definition of the spin current in a coupled bilayer system in
this section. Throughout the whole paper, we consider a coupled
bilayer system where the strengthes of the Rashba-type SOC in each
layer are different and the tunnelling between layers always
occurs. The spaces spanning the electrons' spin states and layer
occupations, respectively, carry out SU(2) representations. If the
spin and layer representations are denoted by Pauli matrices
$\sigma_a$ and $\tau$-matrices $\tau_a$, respectively, the total
Hamiltonian of such a system can be written as
\begin{eqnarray}\label{Hamiltonian}
H_0 \!=\!
 \frac{\hbar^2k^2}{2m}\!+\!
 \left(%
\begin{array}{cc}
  \!\alpha_1\!\! & \!\!0\!\! \\
     \!\!0\!\!   & \!\!\alpha_2\!\! \\
\end{array}%
\right)\!\otimes(k_y\sigma_x \!-\! k_x\sigma_y)
 +\left(%
\begin{array}{cc}
  \!0\!     & \!\beta\! \\
  \!\beta\! & \!0\! \\
\end{array}%
\right)\!\otimes I
  \nonumber \\
 \!\!=\!  \frac{\hbar^2 k^2}{2m}
     \!+\!\Bigl(\alpha_+I \!+\! \alpha_-\tau_z
     \Bigr)\!\otimes(k_y\sigma_x \!-\! k_x\sigma_y)
     \!+\!\beta\tau_x\!\otimes\! I,
\end{eqnarray}
where $\alpha_1$ and $\alpha_2$ refer to SOC strengthes in the
front and back layers, correspondingly, and $\beta$ the tunnelling
strength between layers. $I$ stands for the unit matrix. For
convenience, $\alpha_+ =(\alpha_1+\alpha_2)/2$ and $\alpha_-
=(\alpha_1-\alpha_2)/2$ are introduced in the second line of the
above equation. Hereafter, indices $a$ and $i$ run from 1 to 3.
Let $\psi^{}_{\mathrm f}=(\phi^{}_{\mathrm
f\uparrow},\phi^{}_{\mathrm f\downarrow})^T$ and $\psi^{}_{\mathrm
b}=(\phi^{}_{\mathrm b\uparrow},\phi^{}_{\mathrm b\downarrow})^T$
represent the spin states of the electrons in the front and back
layers, respectively. Hereafter, the layer-index $\mathrm f$ or
$\mathrm b$ labels either the front or back layer. Then a
four-component wave function, denoted by $\Psi=(\phi^{}_{\mathrm
f\uparrow},\phi^{}_{\mathrm f\downarrow}, \phi^{}_{\mathrm
b\uparrow},\phi^{}_{\mathrm b\downarrow})^T =(\psi^{}_\mathrm
f,\psi^{}_\mathrm b)^T$ must be introduced for a complete quantum
mechanical description of the system.

The well accepted definitions of the spin density and the spin
current density in a single-layer system are $S^a \!=\!\Psi^\dag
\sop \Psi$ and $\mathbf J^a \!\!=\! \textrm{Re} \Psi^\dag \js
\Psi$, respectively. Here $\sop =\sigma_a \hbar/2$ is the spin
operator and $\js=\frac{1}{2}\{\mathbf {\hat v}, \sop \}$ the spin
current operator with the curl bracket denoting the
anti-commutator and $\mathbf {\hat v}=\frac{1}{i\hbar}[\mathbf
{\hat r}, H_0]$ the velocity operator. The bold face manifests the
quantity is a vector in the spatial space, \eg $\mathbf
J^a=(J_{x}^a,J_{y}^a,J_{z}^a)$.
It is natural to define the full spin
current operator for the whole bilayer system as
\begin{equation}
\js=\frac{1}{2}\{\mathbf {\hat v}, I\otimes\sop \}\equiv
 \left(%
\begin{array}{cc}
  \js_\mathrm{f} & 0     \\
  0     & \js_\mathrm{b}  \\
\end{array}%
\right),
\end{equation}
with $\js_\mathrm{f}$ and $\js_\mathrm{b}$ being the spin current
operators in the corresponding layers. Even though the tunnelling
couples two layers, the spin current operator is in a block
diagonal form since the tunnelling is momentum-independent. Then
we have the spin density and the spin current density in each
layer
\begin{equation}
S^a_{\ell}= \psi^\dag_{\ell} \!~\sop~\! \psi^{}_{\ell},
  \quad
\mathbf J^a_{\ell} = \textrm{Re}~ \psi^\dag_{\ell}~ \js_{\ell}
~\!\psi^{}_{\ell},
\end{equation}
where $\ell$ stands for $\mathrm{f}$ or $\mathrm{b}$.

It is obvious that the presence of the SOC, which can be regarded
as certain SU(2) gauge potentials $\vec{\mathcal A}_i$ and
$\vec{\mathcal A}_0$~\cite{Li}, leads to the non-conservation of
the spin density. Hereafter, a vector in the spin space is
denoted by an overhead arrow, \eg $\vec S=(S^x, S^y, S^z)$.
In terms of these gauge potentials, the
partially conserved spin current takes a covariant form~\cite{Li}
and obeys the ``continuity-like'' equation, namely,
$\displaystyle\bigl(\deriv{t}-\eta\vec{\mathcal
A_0}\times\bigr)\vec S
 +\bigl(\deriv{x_i}+\eta\vec{\mathcal A}_i\times\bigr)\vec{J}_i
  =0$.
Through an analogous procedure as in Ref.~\cite{Li}, we can derive a
general ``continuity-like'' equation for the spin density in each
single layer in the presence of SU(2) gauge potentials:
\begin{eqnarray}\label{eq:bilayer-Continuity}
&&  \bigl(\deriv{t}
   -\eta\vec{\mathcal A_{\mathrm f 0}}\times\bigr)\sd_\mathrm{f}
   +\bigl(\deriv{x_i}
     +\eta\vec{\mathcal A}_{\mathrm{f} i}
       \times\bigr)\vec{J}^{}_{\mathrm{f} i}
     \hspace{26mm}
  \nonumber \\
&&     \hspace{39mm}
  =\frac{i \beta}{\hbar} (\psi^\dag_{ \mathrm b} \vec s~
                           \psi^{}_{\mathrm f}
    -\psi^\dag_{\mathrm f} \vec s~ \psi^{}_{\mathrm b}),
    \nonumber\\
&&  \bigl(\deriv{t}
  -\eta\vec{\mathcal A_{\mathrm b 0}}\times\bigr)\sd_\mathrm{b}
   +\bigl(\deriv{x_i}
    +\eta\vec{\mathcal A}_{\mathrm{b} i}
      \times\bigr)\vec{J}^{}_{\mathrm{b} i}
         \hspace{26mm}
   \nonumber \\
&&   \hspace{39mm}
   =\frac{i \beta}{\hbar} (\psi^\dag_{ \mathrm f}\vec s~
                            \psi^{}_{\mathrm b}
    -\psi^\dag_{\mathrm b} \vec s~ \psi^{}_{\mathrm f}).
\end{eqnarray}
In the coupled bilayer electron gas with Rashba
SOC, $\vec{\mathcal A}_{\mathrm f x}
\!=\!\frac{2m}{\eta^2}(0,~\alpha_1~,0)$,
$\vec{\mathcal A}_{\mathrm f y}
\!=\!-\frac{2m}{\eta^2}(\alpha_1,~0~
,0)$, $\vec{\mathcal
A}_{\mathrm b x}
\!=\!\frac{2m}{\eta^2}(0,~\alpha_2~,0)$,
$\vec{\mathcal A}_{\mathrm
b y}
\!=\!-\frac{2m}{\eta^2}(\alpha_2,~0~
,0)$ and $\vec{\mathcal
A}_{\mathrm f
z}\!=\!\vec{\mathcal
A}_{\mathrm f 0}\!=\!
\vec{\mathcal A}_{\mathrm b
z}\!=\!\vec{\mathcal
A}_{\mathrm b 0}\!=\!0$
with $\eta=\hbar$.
The tunnelling between layers gives rise to the term on
the right hand side of Eqs.~(\ref{eq:bilayer-Continuity})
and this term results in additional
non-conservations for the spin density in each layer.

\section{Spin currents in a clean system}\label{sec:calculations}

In this section, we calculate the spin currents for a clear system
in Heisenberg
representation~\cite{Shen2003}. A weak electric field $\mathbf E =
E\hat x$ applied on both layers is regarded as a perturbation.
We mainly focus on $J_{\ell\,y}^z$ component of the spin current in
the $\ell$-layer which is flowing perpendicularly to the electric
field with the spin polarized in the $z$-direction.

\begin{figure}[t]
\includegraphics[width=76mm]{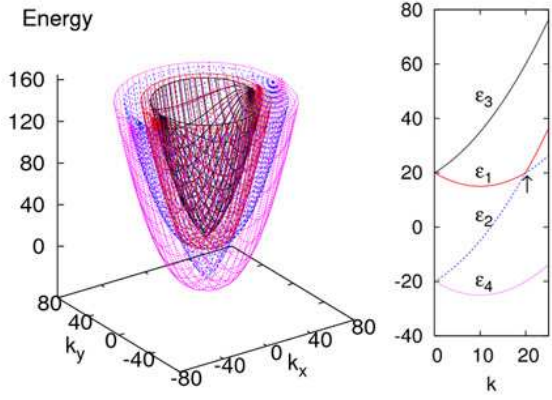}
\caption{(color online) The four energy bands corresponding to the four
eigenstates given in Eq.~(\ref{wavevectors}).
The surface of revolution in the left panel is obtained by revolving the curves
in the right panel with respect to the vertical axis.
The $\uparrow$ in the
right panel marks the level crossing point of $\varepsilon_1$ and
$\varepsilon_2$.
}\label{fig:energyband}
\end{figure}

Diagonalizing the unperturbed Hamiltonian (\ref{Hamiltonian}), we
obtain four energy bands:
\begin{eqnarray}\label{eq:bands}
\varepsilon_1 &=& \frac{\hbar^2k^2}{2m}
        + (\sqrt{\beta^2+\alpha_-^2k^2}-\alpha_+k)~\mathrm{sgn}(k_t-k),
   \nonumber \\
\varepsilon_2 &=& \frac{\hbar^2k^2}{2m}
        - (\sqrt{\beta^2+\alpha_-^2k^2}-\alpha_+k)~\mathrm{sgn}(k_t-k),
    \nonumber \\
\varepsilon_3 &=& \frac{\hbar^2k^2}{2m}
        + \sqrt{\beta^2+\alpha_-^2k^2}+\alpha_+k,
   \nonumber \\
\varepsilon_4 &=& \frac{\hbar^2k^2}{2m}
        - \sqrt{\beta^2+\alpha_-^2k^2}-\alpha_+k,
\end{eqnarray}
with
$$
\mathrm{sgn}(x) = \left \{
\begin{array}{ccc}
1  & ~~\textrm{if} & x>0,  \\
0  & ~~\textrm{if} & x=0,  \\
-1 & ~~\textrm{if} & x<0,
\end{array}
\right.
$$
and $k_t=\beta/\sqrt{\alpha_+^2-\alpha_-^2}$ denoting a special
point where  $\varepsilon_1 = \varepsilon_2 =\hbar^2k^2_t/2m$.
The landscape of these bands are plotted in Fig.~\ref{fig:energyband},
in which $\uparrow$ in the right panel marks the level crossing point
of $\varepsilon_1$ and $\varepsilon_2$ at $k_t$.
As we will see later, the spin Hall
conductivity exhibits sharp cusps around this point.
In the following, we consider the case $k<k_t$
which has the same result as $k>k_t$.
The eigenvectors
$\Psi_{j}=(\psi_{{\mathrm f} j},\psi_{{\mathrm b} j})^T$ with
$j=1,2,3,4$ labelling the band indices are given by
\begin{eqnarray}\label{wavevectors}
\Psi_1 &=& N_1
\left(%
\begin{array}{c}
  i e^{-i\varphi} (\alpha^{}_- k - \!\sqrt{\beta^2+\alpha_-^2k^2}) \\
  - (\alpha^{}_- k - \!\sqrt{\beta^2+\alpha_-^2k^2}) \\
  - i e^{-i\varphi}\beta \\
  \beta \\
\end{array}%
\right),
 \nonumber \\[3mm]
 \Psi_2 &=& N_2
\left(%
\begin{array}{c}
  i e^{-i\varphi} (\alpha^{}_- k - \!\sqrt{\beta^2+\alpha_-^2k^2}) \\
  (\alpha^{}_-k - \!\sqrt{\beta^2+\alpha_-^2k^2}) \\
  i e^{-i\varphi}\beta \\
  \beta \\
\end{array}%
\right),
 \nonumber \\[3mm]
 \Psi_3 &=& N_3
\left(%
\begin{array}{c}
  i e^{-i\varphi}(\alpha^{}_- k \!+\! \sqrt{\beta^2+\alpha_-^2k^2}) \\
  (\alpha^{}_- k \!+\! \sqrt{\beta^2+\alpha_-^2k^2}) \\
  i e^{-i\varphi}\beta \\
  \beta \\
\end{array}%
\right),
 \nonumber \\[3mm]
\Psi_4 &=& N_4
\left(%
\begin{array}{c}
  i e^{-i\varphi}(\alpha^{}_- k  + \!\sqrt{\beta^2+\alpha_-^2k^2}) \\
  -(\alpha^{}_ -k  + \sqrt{\beta^2 + \alpha^2_- k^2}) \\
  -i e^{-i\varphi}\beta \\
  \beta \\
\end{array}%
\right),
\end{eqnarray}
where $\varphi=\tan^{-1}(k_y/k_x)$ and the normalization
coefficients $N_j$ are given in the appendix.

The spin current operator for the whole bilayer system is given by
$\hat{j }_y^z =\displaystyle\frac{1}{2} \{ \hat v_y,I\otimes \sop \}
 =\frac{\hbar^2k_y}{2m} I\otimes\sigma_z$.
Time evolutions of operators are governed by Heisenberg equation
of motion. Thus we have $k_x=\displaystyle
k_{0x}-\frac{eEt}{\hbar}$ and $k_y=k_{0y}$ with $k_{0x}$ and
$k_{0y}$ being the initial values and
\begin{eqnarray}\label{time-evolution}
\frac{\partial}{\partial t}
 \bigl(I\!\otimes\!\sigma_z\bigr) \!&=&\!
 \frac{2}{\hbar}\bigl[\alpha_+ k_x I\otimes\sigma_x
 +\alpha_+k_y I\otimes\sigma_y
 \nonumber \\
&& \hspace{3mm}
 +\alpha_-k_x\tau_z\otimes\sigma_x
 +\alpha_-k_y\tau_z\otimes\sigma_y\bigr].
\end{eqnarray}
Obviously, the time evolution of $I\otimes\sigma_z$ depends on
those of other four-by-four Hermitian matrices, such as
$I\otimes\sigma_x$ which also depends on other matrices. Hence, we
need to deal with the time evolutions of sixteen matrices
$\{I\otimes I, I\otimes \sigma_x, I\otimes \sigma_y, I\otimes
\sigma_z, \tau_x\otimes I, \tau_x\otimes \sigma_x, \tau_x\otimes
\sigma_y, \tau_x\otimes \sigma_z, \tau_y\otimes I, \tau_y\otimes
\sigma_x, \tau_y\otimes \sigma_y, \tau_y\otimes \sigma_z,
\tau_z\otimes I, \tau_z\otimes \sigma_x, \tau_z\otimes \sigma_y,
\tau_z\otimes \sigma_z\}$ which span the space of the four-by-four
Hermitian matrices. If we arrange those 16 matrices successively
in a single column, denoted by $\Gamma$, the problem reduces to
search solutions of a set of 16 linear differential equations:
\begin{equation}\label{diagonalize}
\partial_t \Gamma =\frac{2}{\hbar}
\Bigl(M +\frac{eEt}{\hbar} M_t \Bigr)\Gamma,
\end{equation}
where the concrete expressions of $M$ and $M_t$ are given in the
appendix.

Expanding $\Gamma$ in series of the electric field, namely,
$\Gamma=\Gamma^{(0)}+\Gamma^{(1)} + \cdots$, we have the following
equations:
\begin{eqnarray}
&& \partial_t \Gamma^{(0)}
 =\frac{2}{\hbar}M \Gamma^{(0)},\nonumber \\
&& \partial_t \Gamma^{(1)}
 =\frac{2}{\hbar}M \Gamma^{(1)}+\frac{2eEt}{\hbar^2}M_t\Gamma^{(0)},
\end{eqnarray}
up to the first order. Using the standard method to solve these
equations, we obtain the linear order term
$I\otimes\sigma_z^{(1)}$ in the limit $t\rightarrow 0$:
\begin{eqnarray}\label{Iz}
I\otimes\sigma_z^{(1)}
 = \frac{e E}{2 k (\beta^2+\alpha^2_-k^2-\alpha^2_+k^2)}
    \times\hspace{20mm}
    \nonumber \\
  \Bigl(C_1 I \!\otimes\sigma_{0x} \!+\! C_2 I \!\otimes\sigma_{0y}
    \!+\! C_3\tau_z\!\otimes\sigma_{0x}\!+\!C_4\tau_z\!\otimes\sigma_{0y}
 \Bigr),
\end{eqnarray}
where $\sigma_{0a}$ stand for the initial values of $\sigma_{a}$
at $t=0$ and the coefficients $C$ are written out in the appendix.

The spin currents in both layers produced by the states in each
energy band are evaluated as
\begin{eqnarray}\label{eachJz}
 \langle\psi_{1,\mathrm f}|\hat j^z_y|
  \psi^{}_{1,\mathrm f}\rangle=eE\hbar^2\sin^2\!\!\varphi\times
 \hspace{39mm}
 \nonumber \\
 \frac{
     (\sqrt{\beta^2+\alpha_-^2k^2}-\alpha_-k)
     [\beta^2-(\alpha^2_+-\alpha_+\alpha_-)k^2]}
     {8m\alpha_+k\sqrt{\beta^2+\alpha_-^2 k^2}
     (\beta^2-(\alpha_+^2-\alpha_-^2)k^2)},
  \nonumber \\
 \langle\psi_{3,\mathrm f}|\hat j^z_y|
 \psi^{}_{3,\mathrm f}\rangle =-eE\hbar^2\sin^2\!\!\varphi\times
  \hspace{37mm}
  \nonumber \\
  \frac{
     (\sqrt{\beta^2+\alpha_-^2k^2}+\alpha_-k)
     [\beta^2-(\alpha^2_+-\alpha_+\alpha_-)k^2]}
     {8m\alpha_+k\sqrt{\beta^2+\alpha_-^2 k^2}
     (\beta^2-(\alpha_+^2-\alpha_-^2)k^2)},
  \nonumber \\
 \langle\psi_{1,\mathrm b}|\hat j^z_y|
  \psi^{}_{1,\mathrm b}\rangle =-eE\hbar^2\sin^2\!\!\varphi\times
   \hspace{36mm}
  \nonumber \\
  \frac{
     (\sqrt{\beta^2+\alpha_-^2k^2}+\alpha_-k)
     [(\alpha^2_++\alpha_+\alpha_-)k^2-\beta^2]}
     {8m\alpha_+k\sqrt{\beta^2+\alpha_-^2 k^2}
     (\beta^2-(\alpha_+^2-\alpha_-^2)k^2)},
  \nonumber \\
 \langle\psi_{3,\mathrm b}|\hat j^z_y|
 \psi^{}_{3,\mathrm b}\rangle =eE\hbar^2\sin^2\!\!\varphi\times
  \hspace{39mm}
  \nonumber \\
 \frac{
     (\sqrt{\beta^2+\alpha_-^2k^2}-\alpha_-k)
     [(\alpha^2_++\alpha_+\alpha_-)k^2-\beta^2]}
     {8m\alpha_+k\sqrt{\beta^2+\alpha_-^2 k^2}
     (\beta^2-(\alpha_+^2-\alpha_-^2)k^2)},
 \nonumber
\end{eqnarray}
while
\begin{eqnarray}\label{eq:diff}
\langle\psi_{2,\mathrm f(\mathrm b)}|\hat j^z_y|\psi^{}_{2,\mathrm
f(\mathrm b)}\rangle=-\langle\psi_{1,\mathrm f(\mathrm b)}|\hat
j^z_y|\psi^{}_{1,\mathrm f(\mathrm b)}\rangle,
 \nonumber \\
\langle\psi_{4,\mathrm f(\mathrm b)}|\hat j^z_y|\psi^{}_{4,\mathrm
f(\mathrm b)}\rangle = -\langle\psi_{3,\mathrm f(\mathrm b)}|\hat
j^z_y|\psi^{}_{3,\mathrm f(\mathrm b)}\rangle.
\end{eqnarray}
The total spin current in each layer is the sum of the
contributions of the four bands up to the Fermi level, \ie
$J^z_{\mathrm{f(b)}\,y} =
 \sum_{j,k} \langle \psi_{j,\mathrm {f(b)}}|
  \hat j^z_y|\psi^{}_{j,\mathrm {f(b)}}\rangle
  n^{}_F(\varepsilon_j)/(L_x\times L_y)$
where
$n^{}_F(\varepsilon_j)$ is the Fermi distribution function and
$L_x\times L_y$ the size of the system.
The explicit expressions for $J^z_{\mathrm{f(b)}\,y}$ at zero
temperature are given in the appendix.
Eqs.~(\ref{eq:diff}) tell us that the spin currents produced by
the states in bands $\varepsilon_1$ and $\varepsilon_2$ are always
with the opposite sign. Thus only the contributions by the states
in $\varepsilon_2$ with momentum $k_{F1}<k<k_{F2}$ remain. The
case for bands $\varepsilon_3$ and $\varepsilon_4$ is similar.
Here and throughout the paper, $k_{F j}$ denotes the Fermi wave
vector in the band $\varepsilon_j$.

The full spin current of the whole bilayer system is given by
$J^z_y=J^z_{\mathrm{f}\,y}+J^z_{\mathrm{b}\,y}$ and the
corresponding spin Hall conductivity is defined as
$\sigma_s=\partial{J^z_y}/{\partial E}$.
Our results can also be verified by Kubo formula.
It is worthwhile to observe our results in two specific cases.
In the case that the tunnelling is absent,
the system becomes a decoupled two single
layers and its spin Hall conductivity becomes
$\sigma_s=e/4\pi$, twice of the universal value
in a single layer~\cite{Niu0403}.
In the case of $\alpha_-\rightarrow 0$,
there is no difference between the two
layers and thus they can not be distinguished.
Consequently,
no matter the tunnelling is present or not,
they behave just like decoupled two single layers
since tunnelling to the other layer
makes no difference from staying in the original one.

\begin{figure}[t]
\includegraphics[width=64mm]{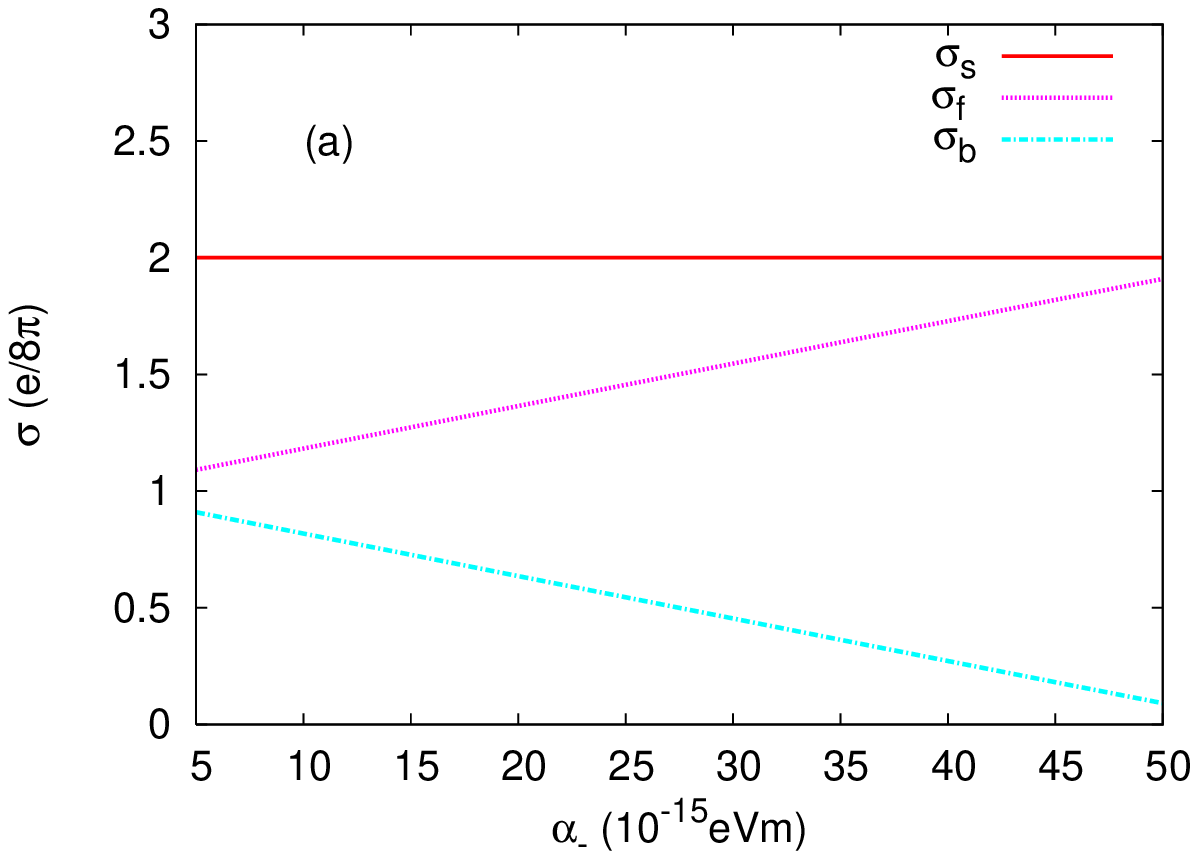}\\[0mm]
\includegraphics[width=42mm]{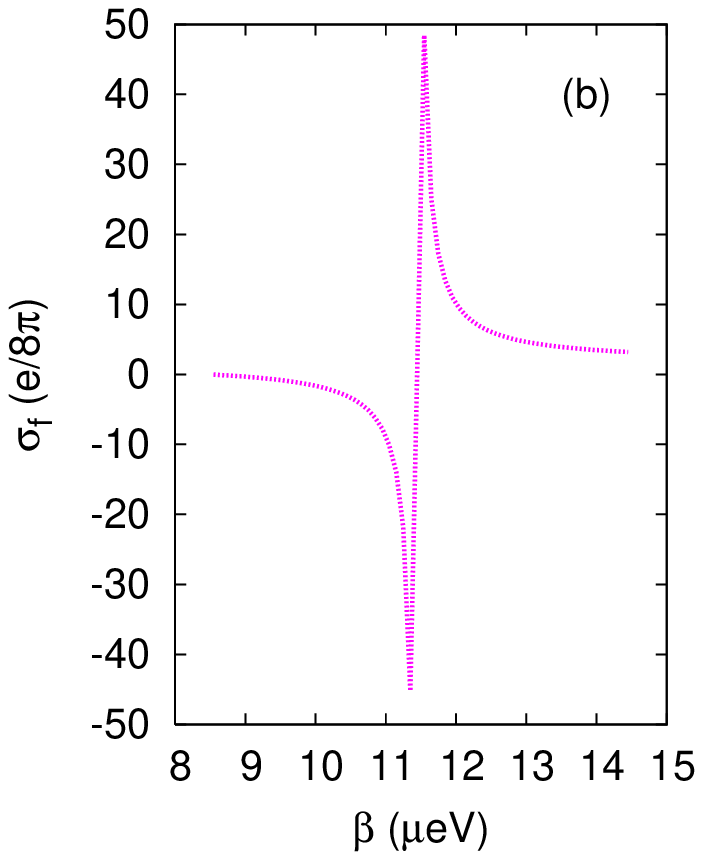}
\includegraphics[width=42mm]{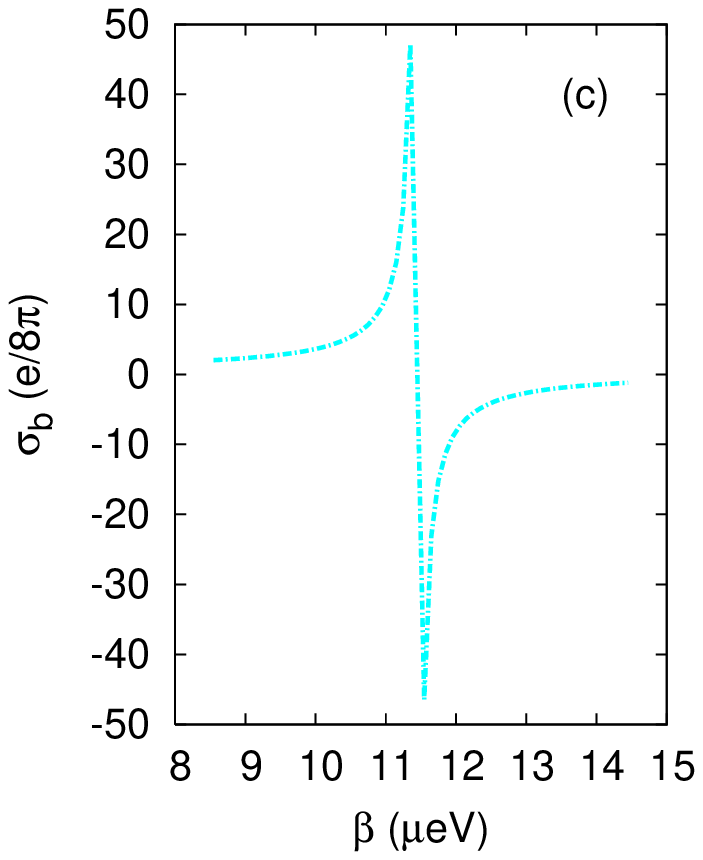}
\caption{(color online)
({\bf a}): Spin currents in each layer and in the whole system versus $\alpha_-$
are plotted with
$\alpha_+=0.55\times10^{-13}\textrm{eVm}$ and
$\beta=4.013\times10^{-4}\textrm{eV}$.
({\bf b}) and ({\bf c}):
Spin currents in each layer versus the tunnelling strength $\beta$
are plotted with
$\alpha_1=10^{-13}\textrm{eVm}$ and
$\alpha_2=10^{-14}\textrm{eVm}$, in which sharp peaks emerge near $\beta_t$.
The choices of the other parameters in the above figures:
the Fermi energy $\varepsilon_F=0.1$eV
and the effective mass $m=0.05m_e$.
}\label{fig:each}
\end{figure}

Now we are in the position to investigate the
tunnelling dependence of spin Hall conductivities $\sigma_\textrm{f}$
and $\sigma_\textrm{b}$
in each layer.
Based on the above results,
we plot $\sigma_\textrm{f}$, $\sigma_\textrm{b}$ as well as
$\sigma_s$ in Fig.~\ref{fig:each}.
The dependence of the spin Hall conductivity in each layer on the strength
of the SOC is quite different from that in a single-layer system
which does not vary as the strength of the SOC changes.
As illustrated in Fig.~\ref{fig:each}(a),
$\sigma^{}_{\textrm{f}(\textrm{b})}$ increases (decreases) monotonously as the
strength of the SOC in its layer increases (decreases) while $\sigma_s$ keeps constant.
We also plot $\sigma_\textrm{f}$ and $\sigma_\textrm{b}$
versus the tunnelling strength in
Fig.~\ref{fig:each}(b-c). They change abruptly near $\beta_t$ where
$k_{F1}=k_{F2}=k_t$ and also undergo sign changes across this point.
At this point, the spin current produced by the states in band
$\varepsilon_1$ and that in band $\varepsilon_2$ cancels each other
precisely, leading to a depression of $\sigma_s$ which always keeps a constant value
$e/4\pi$ for $\beta\neq\beta_t$. It manifests that each layer
posses a large spin conductivity near $\beta_t$ while $\sigma_s$ of the
whole system remains $e/4\pi$. These features are instructive for
designing experiments to detect a magnified spin Hall effect.

\section{The influence of impurities}\label{sec:disorder}

In realistic systems, disorderly distributed impurities are unavoidable,
which frequently affect transport properties.
It was believed that the spin Hall conductivity in a single layer electron gas
could be suppressed to zero by the vertex corrections of nonmagnetic impurities
even for an infinitesimally small
concentration~\cite{Inoue,Halperin,Dimitrova,Raimondi,Rashba}.
Therefore it is necessary
to investigate the impurity effects in the bilayer system.

We consider disorderly deployed nonmagnetic impurities.
The short-ranged interaction between the electron and impurities
at positions $R_i$ is described by
$\hat V_{\textrm{im}}=\sum_{i} u \delta(r-R_i)$.
Here we assume the coupling strength $u$ is sufficiently weak
so that the Born approximation is applicable.
The averaged retarded
Green's function satisfies the Dyson equation
$\bar{G}^R=G^R_0+G^R_0\overline{\Sigma^RG^R}$
where the overline refers to an average taken over
the configuration of impurities,
$G^R_0$ denotes the free Green's function
and $\Sigma^R$ the self energy brought about by the
impurities.
In the Born approximation~\cite{Jauho}, the Dyson
equation can be explicitly written as
\begin{eqnarray}\label{eq:dyson}
\bar{G}^{R}(\vec p,\omega) \!=\! G^{R}_0(\vec p,\omega)
       \hspace{48mm}   \nonumber \\
   +G^{R}_0(\vec p,\omega) \Bigl(u n^{}_\mathrm{im} +\frac{u^2 n^{}_\mathrm{im}}{\nu}
   \sum_{\vec q} \bar{G}^{R}(\vec q,\omega)
   \Bigr)\bar{G}^{R}(\vec p,\omega),
\end{eqnarray}
where $n^{}_\mathrm{im}$ stands for the impurity concentration
and $\nu$ the size of the whole system.

For convenience, we introduce the chiral representation in which the
$H_0$ is diagonalized by a unitary matrix $U$,
\ie
$U^\dag H_0 U=
\textrm{diag}(\varepsilon_1,\varepsilon_2,\varepsilon_3,\varepsilon_4)$.
In this representation,
the free retarded Green's function reads
$G^R_{0(ch)}(\vec p,\omega)
=\textrm{diag}\Bigl((\omega-\varepsilon_1+i\eta)^{-1},
 (\omega-\varepsilon_2+i\eta)^{-1},
 (\omega-\varepsilon_3+i\eta)^{-1},
 (\omega-\varepsilon_4+i\eta)^{-1}\Bigr)$
 such that equation (\ref{eq:dyson}) solves
\begin{equation}
\bar{G}^R_{(\!ch\!)}(\vec p,\omega)
 \!=\! \left(
   \begin{array}{cccc}
     g_1 & 0 & 0 & 0 \\
     0 & g_2 & 0 & 0 \\
     0 & 0 & g_3 & 0 \\
     0 & 0 & 0 & g_4 \\
   \end{array}
 \right),
 \nonumber
\end{equation}
with $g^{}_j=1/(\omega-\varepsilon_j + \frac{i}{2\tau})$ for $j=1,\cdots,4$.
Here
$\tau=(2\pi u^2 n^{}_\mathrm{im}N_F)^{-1}$ is the momentum-relaxation time
and $N^{}_F$ the density of states of the electron at the Fermi surface.

In terms of the Kubo formula, the averaged spin Hall conductivity
at zero temperature can be calculated,
\begin{eqnarray}\label{eq:averaged conductivity}
\bar{\sigma}_s(\omega)=\frac{e}{\omega \nu} \int
  \frac{d\omega'}{2\pi}
  \textrm{Tr}\{\theta(-\omega'-\omega)
          \hspace{29mm}    \nonumber \\
  \times~\overline{\hat j^z_y~ [~G^R(\omega'\!+\!\omega)
  -G^A(\omega'\!+\!\omega)~]~\hat j_e~G^A(\omega')}
         \hspace{7mm}     \nonumber \\
  +~\theta(-\omega')~\overline{\hat j^z_y~G^R(\omega'\!+\!\omega)
  \hat j_e [~G^R(\omega')\!-\!G^A(\omega')~]} \},
      \hspace{1mm}
\end{eqnarray}
where $G^A$ is the advanced Green's function,
$\hat j_e=e\hat v_x$
the charge-current operator
and $\theta(\omega)$ the step function
representing the Fermi distribution function at zero temperature.
The trace $\textrm{Tr}$ in Eq.~(\ref{eq:averaged conductivity})
implies both the conventional trace
over the spin indices and the summation over the momenta.
In the uncrossing approximation~\cite{Dimitrova},
$\bar{\sigma}_s$ is the sum of $\bar{\sigma}^0_s$ and
$\bar{\sigma}^L_s$,
the former is the contribution by one-loop diagram
\begin{center}
\begin{picture}(80,50)(0,-25)
 \put(-2,-2){$\hat j^z_y$}
 \put(15,0){\circle*{8}}
 \qbezier(15,0)(15,17)(40,17)
 \put(37,20){$\bar G^A$}
 \qbezier(40,17)(61,17)(64,4)
 \put(65,0){\circle{8}}
 \put(73,-2){$\hat j_e$,}
 \qbezier(15,0)(15,-17)(40,-17)
 \put(36,-30){$\bar G^R$}
 \qbezier(40,-17)(61,-17)(64,-4)
\end{picture}
\end{center}
while the later is that by a series of ladder diagrams
\begin{center}
\begin{picture}(190,56)(10,-28)
 \put(0,-2){$\hat j^z_y$}
 \put(15,0){\circle*{6}}
 \qbezier(15,0)(15,17)(40,17)
 \put(37,20){$\bar G^A$}
 \qbezier(40,17)(61,17)(64,3)
 \put(65,0){\circle{6}}
 \put(71,-2){$\hat j_e$}
 \qbezier(15,0)(15,-17)(40,-17)
 \put(36,-29){$\bar G^R$}
 \qbezier(40,-17)(61,-17)(64,-3)
 \multiput(26,13)(0,-4){8}{\line(0,1){2}}
 \put(83,-2){$+$}
 \put(96,-2){$\hat j^z_y$}
 \put(110,0){\circle*{6}}
 \qbezier(110,0)(110,17)(135,17)
 \put(131,20){$\bar G^A$}
 \qbezier(135,17)(156,17)(159,3)
 \put(160,0){\circle{6}}
 \put(167,-2){$\hat j_e$}
 \qbezier(110,0)(110,-17)(135,-17)
 \put(130,-29){$\bar G^R$}
 \qbezier(135,-17)(156,-17)(159,-3)
 \multiput(120.5,13)(0,-4){8}{\line(0,1){2}}
 \multiput(133,15)(0,-4){9}{\line(0,1){2}}
 \put(180,-2){$+~\cdots$.}
\end{picture}
\end{center}

\subsection{One-loop diagram contribution}

To derive the dc conductivity, we take the limit
$\omega\rightarrow 0$ in Eq.~(\ref{eq:averaged conductivity})
and obtain the one-loop diagram contribution
\begin{eqnarray}\label{eq:one-loop}
\bar{\sigma}^0_s &=& \frac{e}{2\pi\nu}\textrm{Tr}
   [~\hat j^z_y(\vec p)~\bar G^R(\vec p)
   ~\hat j_e(\vec p)~\bar G^A(\vec p)~]
               \nonumber \\
   &=& \frac{e}{8\pi} \Bigl[\chi^2(1-\frac{1}{1+\Delta^2_{12}\tau^2}
                        +1-\frac{1}{1+\Delta^2_{34}\tau^2})
               \nonumber \\
   && \hspace{6mm} +2(1-\chi^2)(1-\frac{1}{1+\Delta^2_{13}\tau^2})
   \Bigr],
\end{eqnarray}
where
$\chi=\frac{\alpha^{}_- \Delta_{13}}{\alpha^{}_+ \Delta_{14}}$
is a function of $\beta/(\alpha_- k_F)$.
$\Delta_{ij}\equiv\varepsilon_i(k_F)-\varepsilon_j(k_F)$ is the
energy splitting between two bands at the Fermi surface.
The Fermi wave vector $k_F$ is given by $\sqrt{2m\mu}/\hbar$
with $\mu$ being the chemical potential.
In carrying out the summation of momentum in
Eq.(\ref{eq:one-loop}), we have adopted the large
Fermi-circle limit $\mu\gg 1/\tau, \Delta_{ij}$.

Our result in Eq.~(\ref{eq:one-loop}) seems to be similar
to the expression of a single layer system,
$\bar\sigma^0_{sH}=\frac{e}{8\pi}(1-\frac{1}{1+\Delta^2\tau^2})$.
Moreover, the extra term in the last line of Eq.(\ref{eq:one-loop})
and the pre-factors $\chi^2$ are peculiar in the bilayer system.
It is worthwhile to observe the aforementioned two specific cases.
In the case of zero tunnelling $\beta\rightarrow 0$ ($\chi=-1$),
we have $\bar{\sigma}^0_s = \frac{e}{8\pi}
( 1-\frac{1}{1+\Delta^2_{f}\tau^2}
+1-\frac{1}{1+\Delta^2_{b}\tau^2} )$ where
$\Delta_{f}=2\alpha_1k_F$ and $\Delta_{b}=2\alpha_2k_F$ are the
spin-orbit splittings in each layer.
It demonstrates that the system reduces to a decoupled one.
In the twin-layer case $\alpha_-\rightarrow 0$ ($\chi=0$),
we have $\bar{\sigma}^0_s =
\frac{e}{4\pi} (1-\frac{1}{1+\Delta^2\tau^2} )$
which is just twice of the value of a single layer system.
This is actually a trivial case
as there is no difference between layers
even though the tunnelling is present.
The above reasonable conclusions are consistent
with the results for the clean system
derived in previous section.

\subsection{Vertex correction}

The sum of ladder diagrams gives rise to $\bar\sigma^L_s$.
By introducing a matrix-valued vertex
$\tilde J^z_y$ which is the sum of the vertex correction
to $\hat j^z_y$, diagrammatically,
\begin{center}
\begin{picture}(130,30)(15,-15)
 \put(0,-2){$\tilde J^z_y ~\equiv~$}
 \put(30,-2){$\hat j^z_y$}
 \put(45,0){\circle*{6}}
 \qbezier(45,0)(45,11)(57,13)
 \qbezier(45,0)(45,-11)(57,-13)
 \multiput(57,11)(0,-4){7}{\line(0,1){2}}
 \put(67,-2){$+$}
 \put(80,-2){$\hat j^z_y$}
 \put(95,0){\circle*{6}}
 \qbezier(95,0)(95,15)(118,15)
 \qbezier(95,0)(95,-15)(118,-15)
 \multiput(106,11)(0,-4){7}{\line(0,1){2}}
 \multiput(118,13)(0,-4){8}{\line(0,1){2}}
 \put(130,-2){$+~\cdots$}
\end{picture}
\end{center}
then $\bar\sigma^L_s$ can be written as
\begin{eqnarray}
\bar\sigma^L_s=\frac{e}{2\pi\nu}\textrm{Tr}
   [~\tilde J^z_y~\bar G^R(\vec p)
   ~\hat j_e(\vec p)~\bar G^A(\vec p)~],
\end{eqnarray}
where $\tilde J^z_y$ is momentum-independent and satisfies the
transfer matrix equation
\begin{eqnarray}\label{eq:self-consis for vertex}
\tilde J^z_y=\frac{u^2 n^{}_{\textrm{im}}}{\nu}\sum_{\vec q}
   \bar G^A(\vec q)(\hat j^z_y(\vec q)+\tilde J^z_y)\bar G^R(\vec q).
\end{eqnarray}
As a result, we have
\begin{widetext}
\begin{eqnarray}
\bar\sigma^L_s=\frac{e}{8\pi}\frac{\Delta_{13}\tau}{v_F}
     \Bigl\{ (2-\frac{\chi^2+\alpha_-\chi/\alpha_+}{1+\Delta^2_{12}\tau^2}
            -\frac{\chi^2-\alpha_+\chi/\alpha_+}{1+\Delta^2_{34}\tau^2}
            -\frac{2(1-\chi^2)}{1+\Delta^2_{13}\tau^2})~i(J_{12}+J_{34})
                  \hspace{46mm}      \nonumber \\
      +\frac{\alpha_-}{\alpha_+}
        (2-\frac{1+\alpha_+\chi/\alpha_-}{1+\Delta^2_{12}\tau^2}
          -\frac{1-\alpha_+\chi/\alpha_-}{1+\Delta^2_{34}\tau^2}
          -\frac{2(1-\chi^2)}{1+\Delta^2_{14}\tau^2})~i(J_{12}-J_{34})
                  \hspace{43mm}       \nonumber \\
      -\frac{2\alpha_-\sqrt{1-\chi^2}\tau}{\alpha_+}
       [\frac{\Delta_{12}}{1+\Delta^2_{12}\tau^2}(1+\frac{\Delta_{13}}{\Delta_{14}})
       +\frac{\Delta_{34}}{1+\Delta^2_{34}\tau^2}(1-\frac{\Delta_{13}}{\Delta_{14}})
       +\frac{2\Delta_{14}}{1+\Delta^2_{14}\tau^2}
       -\frac{\Delta_{13}}{\Delta_{14}}
         \frac{2\Delta_{13}}{1+\Delta^2_{13}\tau^2}]J_{14}  \Bigr\}.
\end{eqnarray}
\end{widetext}
where $J_{12},J_{14},J_{34}$ are the matrix elements of $\tilde J^z_y$.
Their explicit expressions, a solution of
Eq.~(\ref{eq:self-consis for vertex}),
are given in Eq.~(\ref{eq:J1234}) in the appendix.
When the tunnelling vanishes, we have
$\bar\sigma^L_s=-\frac{e}{8\pi}
(1-\frac{1}{1+\Delta^2_{f}\tau^2}
+1-\frac{1}{1+\Delta^2_{b}\tau^2}) $, reducing to the case
of a decoupled bilayer system, and $\bar\sigma^L_s$ precisely
cancels $\bar\sigma^0_s$, leading to a vanishing spin Hall
conductivity. The nontrivial situation is that both the tunnelling
$\beta$ and the difference in Rashba strengthes $\alpha_-$ are
present, which makes $\bar\sigma_s$ survives.

\section{Magnification effect and possible experiments }
\label{sec:magnification}

\begin{figure}[h]
\includegraphics[width=76mm]{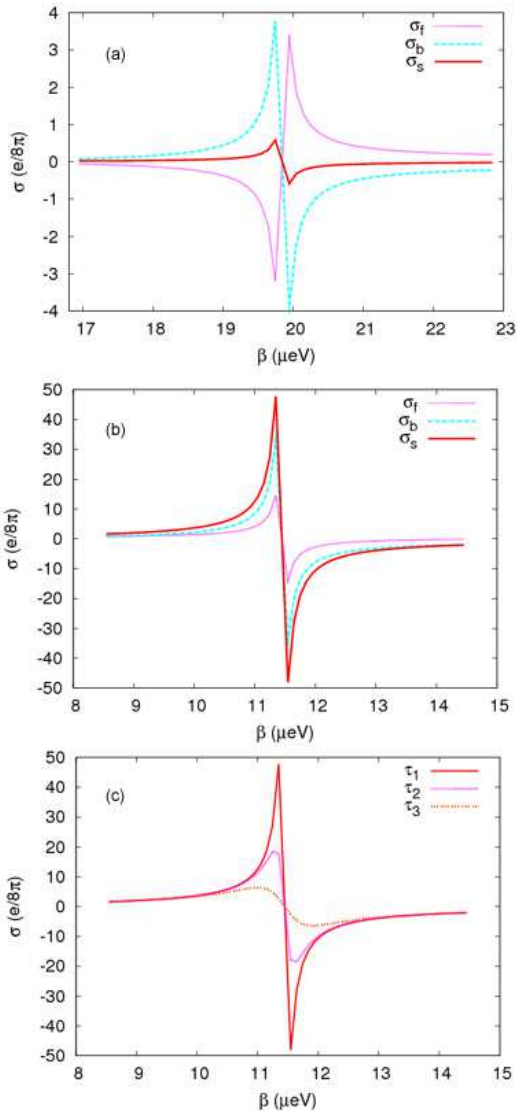}
\caption{(color online) Spin conductivities in each layer
$\bar\sigma_{\mathrm {f(\!b\!)}}$ and the whole system
$\bar\sigma_s$ are plotted with parameters $\tau=6.6\textrm{ns}$,
$\alpha_+=0.55\times10^{-13}\textrm{eVm}$, and
$\alpha_- = 0.45\times10^{-14}\textrm{eVm}$ in panel (a) while
$\alpha_- = 0.45\times10^{-13}\textrm{eVm}$ in panel (b).
Clearly, sharp cups show up around the turning point $\beta_t$.
Panel (c) is the plot of $\bar\sigma_s $ with different momentum
relaxation times:
$\tau_1=6.6\textrm{ns}, \tau_2= 2.1\textrm{ns}, \tau_3=0.66\textrm{ns}$
while the other parameters are the same as in panel (b).
}\label{fig:results}
\end{figure}

The spin Hall conductivity is the sum of $\bar\sigma^0_s$ and
$\bar\sigma^L_s$. An arbitrarily small concentration of nonmagnetic
impurities can not suppress the spin Hall conductivity in a bilayer
system to zero, which is quite different from the case in the single layer
system.
In Fig.~(\ref{fig:results}), we plot the spin Hall conductivities
for each layer $\bar\sigma_{\mathrm {f(\!b\!)}}$ and for the
whole system $\bar\sigma_s$ with different parameters.
Panel (a) of Fig.~(\ref{fig:results}) is the plot for
$\alpha_+=0.55\times10^{-13}\textrm{eVm}$ and
$\alpha_-=0.45\times10^{-14}\textrm{eVm}$ (\ie The strengthes of the
Rashba spin-orbit coupling in each layer are of the same order).
The curves for the conductivities exhibit similar cusps around
the turning point $\beta_t$ as in
Fig.~(\ref{fig:each}) without impurities.
The conductivity in each
layer possess opposite signs, leading to a quite small
$\bar\sigma_s$ for the whole system.
However, things are changed when $\alpha_-$ is comparably large.
Panel (b) in Fig.~(\ref{fig:results}) shows the conductivities with parameters
$\alpha_+=0.55\times10^{-13}\textrm{eVm}$ and
$\alpha_-=0.45\times10^{-13}\textrm{eVm}$, \ie the Rashba strength
in the front layer is ten times as much as that in the back layer.
The opposite signs of the conductivities in each layer
in the absence of impurities
turn to be the same in presence of impurities.
As a result,
the peak value of $\bar\sigma_s$ for the whole system is
considerably large. It suggests that a large difference in the
strength of Rashba spin-orbit coupling between layers is favorable
for a greatly enhanced spin Hall conductivity.

Above results are obtained with rather dilute impurities which
requires the mobility of the two-dimensional electron gas to be
quite high. The influence of the concentration of impurities on the
spin Hall conductivities is also studied, as shown in panel (c) in
Fig.~(\ref{fig:results}). Increasing the concentration of impurities, we
find that the peak value decreases. Although the impurities
tend to suppress the spin Hall conductivity, $\bar\sigma_s$ would
still be detectable. For a two-dimensional electron gas with its
mobility of order $10^6 \textrm{cm}^2/\textrm{Vs}$ which is in an
experimentally realizable regime, the peak value of
$\bar\sigma_s$ is around $e/8\pi$. Thus the spin Hall conductivity
for a bilayer electron system does not vanish and is expected to be
measured in samples with high mobility.

\begin{figure}[h]
\includegraphics[width=58mm]{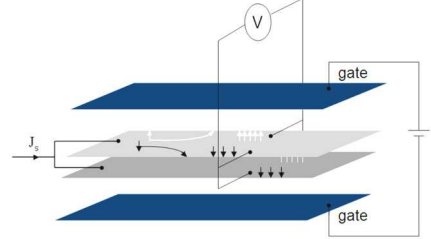}
\caption{(color online) A proposed experimental scheme to detect
a magnification of the spin Hall effect.
By injecting spin current (charge current with spin mostly polarized)
into a bilayer spin Hall bar and connecting the voltmeter
to measure the transverse voltage,
a greatly enhanced Hall voltage
is expected to be observed near the turning point $\beta_t$.}
\label{fig:illu}
\end{figure}

We discuss possible experiments to detect the magnification
of the spin Hall effect.
Our proposal is based on the fact that a spin-polarized electric current
(means the existence of spin current)
in the presence of the SOC can induce different charge populations at
the laterals and hence a Hall voltage can be
detected~\cite{Valenzuela}.
Since the induced Hall voltage is in proportional to the spin Hall conductivity,
its magnitude is greatly enhanced near the turning point $\beta_t$
in the coupled bilayer electron gas. As the tunnelling strength can be tuned
by the gate voltage, we therefore suggest experimentally detect an enormously
magnified Hall voltage by tuning the
tunnelling strength to be near $\beta_t$ in the bilayer system
(see Fig.~\ref{fig:illu}).
Additionally, the sign changes of
spin Hall conductivity across $\beta_t$ also
make the coupled bilayer system a candidate for fabricating possible
logical gates.

\section{Summary}\label{sec:summary}

We investigated the properties of the spin transport in a coupled
bilayer system where the strength of the SOC in each layer may be
different and the tunnelling between the two layers occurs. We gave
natural definitions of the spin density and the spin current density
in each layer and derived the corresponding ``continuity-like''
equations. Based on the calculations in Heisenberg representation,
we obtained the spin current. The curves of the spin Hall conductivities
in each layer exhibit sharp cusps
around the turning point and the peak values have signs changed across this
point. We also investigated the influence of impurities on the spin Hall
conductivity. We found that an arbitrarily small concentration of nonmagnetic
impurities do not suppress the spin Hall conductivity to zero in a bilayer
system, which is quite different from the case in the single layer
system. The opposite signs of the conductivities in the absence of impurities
become the same in presence of impurities.
Making use of these features, we proposed a possible
experiment to detect a magnified spin Hall effect by direct
electronic measurement.
The sign-change property may also be used in
designing certain logical gates.

\acknowledgements
The work was supported by Program for Changjiang Scholars and Innovative
Research Team in University, and NSFC Grant Nos. 10225419 and 10674117.

\appendix
\section{Expressions for some coefficients and the matrices}
\begin{widetext}
The coefficient matrices in the linear Eqs.~(\ref{diagonalize})
are written as
\begin{eqnarray}
M=\left(
\small
\begin{array}{cccccccccccccccc}
  0 &  0   &  0   &  0   &  0   &  0   &  0   &  0   &  0   &  0   &  0   &  0   &    0  &  0   &  0   &  0    \\
  0 &  0   &  0   &-f_x^+&  0   &  0   &  0   &  0   &  0   &  0   &  0   &  0   &    0  &  0   &  0   &-f_x^- \\
  0 &  0   &  0   &-f_y^+&  0   &  0   &  0   &  0   &  0   &  0   &  0   &  0   &    0  &  0   &  0   &-f_y^- \\
  0 & f_x^+& f_y^+&  0   &  0   &  0   &  0   &  0   &  0   &  0   &  0   &  0   &    0  & f_x^-& f_y^-&  0   \\
  0 &  0   &  0   &  0   &  0   &  0   &  0   &  0   &  0   &-f_y^-& f_x^-&  0   &    0  &  0   &  0   &  0   \\
  0 &  0   &  0   &  0   &  0   &  0   &  0   &-f_x^+&-f_y^-&  0   &  0   &  0   &    0  &  0   &  0   &  0   \\
  0 &  0   &  0   &  0   &  0   &  0   &  0   &-f_y^+& f_x^-&  0   &  0   &  0   &    0  &  0   &  0   &  0   \\
  0 &  0   &  0   &  0   &  0   & f_x^+& f_y^+&  0   &  0   &  0   &  0   &  0   &    0  &  0   &  0   &  0   \\
  0 &  0   &  0   &  0   &  0   & f_y^-&-f_x^-&  0   &  0   &  0   &  0   &  0   & -\beta&  0   &  0   &  0 \\
  0 &  0   &  0   &  0   & f_y^-&  0   &  0   &  0   &  0   &  0   &  0   &-f_x^+&    0  &-\beta&  0   &  0 \\
  0 &  0   &  0   &  0   &-f_x^-&  0   &  0   &  0   &  0   &  0   &  0   &-f_y^+&    0  &  0   &-\beta&  0 \\
  0 &  0   &  0   &  0   &  0   &  0   &  0   &  0   &  0   & f_x^+& f_y^+&  0   &    0  &  0   &  0   &-\beta \\
  0 &  0   &  0   &  0   &  0   &  0   &  0   &  0   & \beta&  0   &  0   &  0   &    0  &  0   &  0   &  0 \\
  0 &  0   &  0   &-f_x^-&  0   &  0   &  0   &  0   &  0   & \beta&  0   &  0   &    0  &  0   &  0   &-f_x^+ \\
  0 &  0   &  0   &-f_y^-&  0   &  0   &  0   &  0   &  0   &  0   & \beta&  0   &    0  &  0   &  0   &-f_y^+ \\
  0 & f_x^-& f_y^-&  0   &  0   &  0   &  0   &  0   &  0   &  0   &  0   & \beta&    0  & f_x^+& f_y^+&  0 \\
\end{array}%
\right),
\end{eqnarray}
where $f_{x}^{\pm}=\alpha_{\pm} k_{0x}$ and
$f_{y}^{\pm}=\alpha_{\pm} k_{0y}$, and
\begin{eqnarray}
M_t=\left(%
\begin{array}{cccccccccccccccc}
  0 &  0      &  0 &  0      &  0     &     0   &  0     &  0     &  0      &  0      &  0      &  0     & 0 &     0   &  0   &  0     \\
  0 &  0      &  0 & \alpha_+&  0     &     0   &  0     &  0     &  0      &  0      &  0      &  0     & 0 &     0   &  0   &\alpha_- \\
  0 &  0      &  0 &  0      &  0     &     0   &  0     &  0     &  0      &  0      &  0      &  0     & 0 &     0   &  0   &  0 \\
  0 &-\alpha_+&  0 &  0      &  0     &     0   &  0     &  0     &  0      &  0      &  0      &  0     & 0 &-\alpha_-&  0   &  0   \\
  0 &  0      &  0 &  0      &  0     &     0   &  0     &  0     &  0      &  0      &-\alpha_-&  0     & 0 &     0   &  0   &  0   \\
  0 &  0      &  0 &  0      &  0     &     0   &  0     &\alpha_+&  0      &  0      &  0      &  0     & 0 &     0   &  0   &  0   \\
  0 &  0      &  0 &  0      &  0     &     0   &  0     &  0     &-\alpha_-&  0      &  0      &  0     & 0 &     0   &  0   &  0   \\
  0 &  0      &  0 &  0      &  0     &-\alpha_+&  0     &  0     &  0      &  0      &  0      &  0     & 0 &     0   &  0   &  0   \\
  0 &  0      &  0 &  0      &  0     &     0   &\alpha_-&  0     &  0      &  0      &  0      &  0     & 0 &     0   &  0   &  0 \\
  0 &  0      &  0 &  0      &  0     &     0   &  0     &  0     &  0      &  0      &  0      &\alpha_+& 0 &     0   &  0   &  0 \\
  0 &  0      &  0 &  0      &\alpha_-&     0   &  0     &  0     &  0      &  0      &  0      &  0     & 0 &     0   &  0   &  0 \\
  0 &  0      &  0 &  0      &  0     &     0   &  0     &  0     &  0      &-\alpha_+&  0      &  0     & 0 &     0   &  0   &  0 \\
  0 &  0      &  0 &  0      &  0     &     0   &  0     &  0     &  0      &  0      &  0      &  0     & 0 &     0   &  0   &  0 \\
  0 &  0      &  0 & \alpha_-&  0     &     0   &  0     &  0     &  0      &  0      &  0      &  0     & 0 &     0   &  0   &\alpha_+ \\
  0 &  0      &  0 &  0      &  0     &     0   &  0     &  0     &  0      &  0      &  0      &  0     & 0 &     0   &  0   &  0 \\
  0 &-\alpha_-&  0 &  0      &  0     &     0   &  0     &  0     &  0      &  0      &  0      &  0     & 0 &-\alpha_+&  0&  0    \\
\end{array}%
\right).
\end{eqnarray}
The normalization coefficients for the eigenvectors in
Eqs.~(\ref{wavevectors}) read
\begin{eqnarray}
N_1 &=& \frac{1}{2} \Bigl[\beta^2+\alpha^2_- k^2- \alpha_-k
\sqrt{\beta^2+\alpha_-^2 k^2}~\Bigr]^{-1/2},
  \nonumber \\
 N_2 &=& \frac{1}{2\beta}\Bigl[1
 +\alpha^{}_- k/\sqrt{\beta^2+\alpha_-^2k^2}~ \Bigr]^{1/2},
 \nonumber \\
 N_3 &=& \frac{1}{2} \Bigl[\beta^2+\alpha^2_-k^2+\alpha_-k
 \sqrt{\beta^2+\alpha_-^2k^2}~\Bigr]^{-1/2},
  \nonumber \\
 N_4 &=& \frac{1}{2\beta}\Bigl[1
      -\alpha^{}_- k/\sqrt{\beta^2+\alpha_-^2k^2} \Bigr]^{1/2}.
\end{eqnarray}
The coefficients in the Eq.~(\ref{Iz}) are written as
\begin{eqnarray}
 C_1 &=& -\frac{(\beta^2-\alpha^2_+k^2)\sin^2\!\varphi}{\alpha_+k},
 \nonumber \\
 C_2 &=&\frac{(\beta^2-\alpha^2_+k^2)\sin\!\varphi\cos\!\varphi}
 {\alpha_+k},
  \nonumber \\
 C_3 &=& \frac{\alpha_-k}{2(\beta^2+\alpha^2_-k^2)^2}
 \Bigl[-2\beta^4-2\sin^2\!\varphi~\alpha^4_-k^4
   \nonumber \\
   && + \beta^2(2\cos^2\!\varphi~\alpha_+^2
       +(\cos\!2\varphi-3)~\alpha^2_-)k^2
       \Bigr],
  \nonumber \\
 C_4 &=& \frac{\alpha_-k^3\sin\!2\varphi}
   {2(\beta^2+\alpha^2_-k^2)^2}
 \Bigl[\beta^2(\alpha_-^2+\alpha^2_+)+\alpha^4_-k^2\Bigr].
\end{eqnarray}
The total spin current for each layer in a realistic sample of
size $L_x\times L_y$ is given by
\begin{eqnarray*}
J^z_{\mathrm{f}\,y} =
  \frac{1}{L_xL_y}\sum_{i=1}^4\sum_k\psi_{i,\mathrm f}^\dag~
  \hat j^z_y~\psi^{}_{i,\mathrm f}
  = \frac{\hbar^2 e E}{32\pi m\alpha_+(\alpha_+\!+\alpha_-)}\times
  \hspace{85mm}
  \nonumber \\
  \left\{~\Bigl[\alpha_+k
         \!-\! \frac{\alpha_+}{\alpha_-}\sqrt{\beta^2\!+\!\alpha_-^2k^2}
                     \right.
         \!-\! \frac{\alpha_-\beta}{2\sqrt{\alpha_+^2\!-\!\alpha_-^2}}
           \Bigl(\ln\bigl|\frac{\sqrt{\alpha_+^2\!-\!\alpha_-^2}k\!-\!\beta}
         {\sqrt{\alpha_+^2 \!-\!\alpha_-^2}k \!+\!\beta}\bigr|
  -\frac{\alpha_-}{\alpha_+}
  \ln\bigl|\frac{\sqrt{\alpha_+^2\!-\!\alpha_-^2}
   \sqrt{\beta^2\!+\!\alpha_-^2k^2}-\alpha_+\beta}
   {\sqrt{\alpha_+^2\!-\!\alpha_-^2}
   \sqrt{\beta^2\!+\!\alpha_-^2k^2}\!+\!\alpha_+\beta}
   \bigr|\Bigr)~\Bigr]^{k_{F1}}_{k_{F2}}
        \nonumber \\
 -\Bigl[\alpha_+k
         \!+\!\frac{\alpha_+}{\alpha_-}\sqrt{\beta^2\!+\!\alpha_-^2k^2}
         \!-\!\frac{\alpha_-\beta}{2\sqrt{\alpha_+^2\!-\!\alpha_-^2}}
           \Bigl(\ln\bigl|\frac{\sqrt{\alpha_+^2\!-\!\alpha_-^2}k-\beta}
         {\sqrt{\alpha_+^2\!-\!\alpha_-^2}k\!+\!\beta}\bigr|
 \!+\!\left.
  \frac{\alpha_-}{\alpha_+}
  \ln\bigl|\frac{\sqrt{\alpha_+^2\!-\!\alpha_-^2}
  \sqrt{\beta^2\!+\!\alpha_-^2k^2}\!-\!\alpha_+\beta}
   {\sqrt{\alpha_+^2\!-\!\alpha_-^2}
   \sqrt{\beta^2\!+\!\alpha_-^2k^2}\!+\!\alpha_+\beta}
   \bigr| \Bigr) \Bigr]^{k_{F3}}_{k_{F4}}
   \right\},
\end{eqnarray*}
\begin{eqnarray}\label{totJz}
J^z_{\mathrm{b}\,y} =
   \frac{1}{L_xL_y}\sum_{i=1}^4\sum_k\psi_{i,\mathrm b}^\dag~
   \hat j^z_y~\psi^{}_{i,\mathrm b}
 = \frac{\hbar^2 e E}{32\pi m\alpha_+(\alpha_+\!-\alpha_-)}\times
  \hspace{85mm}
      \nonumber \\
 \left\{ \Bigl[\alpha_+k
         \!+\! \frac{\alpha_+}{\alpha_-}\sqrt{\beta^2 \!+\! \alpha_-^2k^2}
         \!+\! \frac{\alpha_-\beta}{2\sqrt{\alpha_+^2\!-\!\alpha_-^2}}
           \Bigl(\ln\bigl|\frac{\sqrt{\alpha_+^2\!-\!\alpha_-^2}k \!-\!\beta}
         {\sqrt{\alpha_+^2\!-\!\alpha_-^2}k \!+\!\beta}\bigr|
     \right.
 +\frac{\alpha_-}{\alpha_+}
   \ln\bigl|\frac{\sqrt{\alpha_+^2-\alpha_-^2}\sqrt{\beta^2+\alpha_-^2k^2}-\alpha_+\beta}
   {\sqrt{\alpha_+^2-\alpha_-^2}\sqrt{\beta^2+\alpha_-^2k^2}+\alpha_+\beta}
   \bigr|\Bigr)\Bigr]^{k_{F1}}_{k_{F2}}
  \nonumber \\
 -\Bigl[\alpha_+k
         \!-\! \frac{\alpha_+}{\alpha_-}\sqrt{\beta^2 \!+\! \alpha_-^2k^2}
        \! +\! \frac{\alpha_-\beta}{2\sqrt{\alpha_+^2 \!-\! \alpha_-^2}}
           \Bigl(\ln\bigl|\frac{\sqrt{\alpha_+^2 \!-\! \alpha_-^2}k \!-\!\beta}
                               {\sqrt{\alpha_+^2 \!-\! \alpha_-^2}k \!+\! \beta}\bigr|
 \left. -\frac{\alpha_-}{\alpha_+}
  \ln\bigl|\frac{\sqrt{\alpha_+^2 \!-\! \alpha_-^2}
                \sqrt{\beta^2 \!+\! \alpha_-^2k^2}\!-\!\alpha_+\beta}
                {\sqrt{\alpha_+^2 \!-\! \alpha_-^2}
                \sqrt{\beta^2 \!+\! \alpha_-^2k^2}\!+\!\alpha_+\beta}
   \bigr|\Bigr)\Bigr]^{k_{F3}}_{k_{F4}}
   \right\}.
\end{eqnarray}
The above results are derived by assuming that the special point
$k_t$ is far away from the Fermi momenta. When
$k_{F3}<k_t=k_{F1}=k_{F2}<k_{F4}$, the spin current produced by
the state with $k_t$ is given by
$\hbar^2(2\alpha_+^2-\alpha_-^2)/4m\alpha_+^3k_t$ in unit of
$eE/4\pi$.

The matrix elements $J_{12}, J_{14}, J_{34}$ of $\tilde J^Z_y$ can
be solved from the Eq.(\ref{eq:self-consis for vertex}), which
is in fact a task of solving a set of linear equations
\end{widetext}
\begin{equation}
\left(
  \begin{array}{ccc}
    w_{+}       &  2\lambda_{+} &  \tilde w \\
    \lambda_{+} &  w_{m}        & -\lambda_{-} \\
    \tilde w    & -2\lambda_{-} &  w_{-}  \\
  \end{array}
\right)
\left(
  \begin{array}{c}
    J_{12} \\
    J_{14} \\
    J_{34} \\
  \end{array}
\right) =
\left(
  \begin{array}{c}
    -q_{+} \\
     q_{m} \\
     q_{-} \\
  \end{array}
\right).
\end{equation}
The solutions are given by
\begin{eqnarray}\label{eq:J1234}
J_{12}=\frac{1}{d} [~q_{+}(w_{-}w_{m} - 2\lambda_{-}^2)
        +2q_{m}(w_-\lambda_{+} + \tilde w\lambda_{-})
                 \hspace{9mm}  \nonumber \\
        -q_{-} (\tilde w w_{m} + 2\lambda_{+}\lambda_{-})~],
                 \hspace{34mm}  \nonumber \\
J_{14}=-\frac{1}{d}[~q_{+}(w_{-}\lambda_{+}
        + \tilde w \lambda_{-})
        + q_{m} (w_{+} w_{-} -\tilde w^2)
                 \hspace{11mm}  \nonumber \\
        -q_{-} (\tilde w\lambda_{+}+w_{+}\lambda_{-})~],
                 \hspace{35mm}  \nonumber \\
J_{34}=-\frac{1}{d}[~q_{+} (\tilde w w_{m}
        + 2\lambda_{+}\lambda_{-})
        + 2q_{m} (w_{+}\lambda_{-}
        + \tilde w\lambda_{+})
                 \hspace{5mm}  \nonumber \\
        -q_- (w_{+}w_m-2\lambda_{+}^2)~],
                 \hspace{36mm}
\end{eqnarray}
with coefficients
\[
d \!=\! (\tilde w^2 \!-\! w_{+}w_{-})w_{m}
   \!+\! 2(w_{-}\lambda_{+}^2 \!+\! w_{+}\lambda_{-}^2
   \!+\! 2\tilde w\lambda_{+}\lambda_{-}),
\]
\begin{eqnarray}
w_{\pm} = \frac{1}{4}\Bigl[2
   +(1-\chi^2)(1-\frac{1}{1+\Delta^2_{13}\tau^2}
   -\frac{1}{1+\Delta^2_{14}\tau^2})
               \nonumber \\
   -\frac{\half(\chi^2+1)\pm\chi}{1+\Delta^2_{12}\tau^2}
   -\frac{\half(\chi^2+1)\mp\chi}{1+\Delta^2_{34}\tau^2} \Bigr],
               \hspace{14mm} \nonumber \\
w_{m} = 1-\frac{1}{2(1+\Delta^2_{14}\tau^2)} -\frac{1}{4}
   \Bigl[\frac{2\chi^2}{1+\Delta^2_{13}\tau^2}
               \hspace{16mm} \nonumber \\
   +(1-\chi^2)(\frac{1}{1+\Delta^2_{12}\tau^2}
   +\frac{1}{1+\Delta^2_{34}\tau^2}) \Bigr],
               \hspace{10mm} \nonumber \\
\tilde w = \frac{1-\chi^2}{8}
   \Bigl[\frac{1}{1+\Delta^2_{12}\tau^2}
        +\frac{1}{1+\Delta^2_{34}\tau^2}
               \hspace{21mm} \nonumber \\
   -2(1+\frac{1}{1+\Delta^2_{13}\tau^2}
       -\frac{1}{1+\Delta^2_{14}\tau^2})  \Bigr],
               \hspace{14mm} \nonumber \\
\lambda_{\pm} = \frac{i\sqrt{1-\chi^2}\tau}{8}
  \Bigl[\frac{\Delta_{12}(1\pm\chi)}{1+\Delta^2_{12}\tau^2}
  +\frac{\Delta_{34}(1\mp\chi)}{1+\Delta^2_{34}\tau^2}
               \hspace{10mm} \nonumber \\
  \mp\frac{\chi\Delta_{13}}{1+\Delta^2_{13}\tau^2}
  +\frac{\Delta_{14}}{1+\Delta^2_{14}\tau^2}  \Bigr],
               \hspace{25mm} \nonumber \\
q_{\pm} = -\frac{iv_F\tau}{8}
  \Bigl[\frac{\Delta_{12}(\chi^2\pm\chi)}{1+\Delta^2_{12}\tau^2}
  -\frac{\Delta_{34}(\chi^2\mp\chi)}{1+\Delta^2_{34}\tau^2}
               \hspace{12mm} \nonumber \\
  +\frac{2(1-\chi^2)\Delta_{13}}{1+\Delta^2_{13}\tau^2} \Bigr],
               \hspace{42mm} \nonumber \\
q_{m} = -\frac{v_F\chi\sqrt{1-\chi^2}}{8}
  (\frac{1}{1+\Delta^2_{12}\tau^2}
  +\frac{1}{1+\Delta^2_{34}\tau^2}
               \hspace{7mm} \nonumber \\
  -\frac{2}{1+\Delta^2_{13}\tau^2}).
               \hspace{48mm}
\end{eqnarray}
where $v_F$ is the Fermi velocity.

\end{document}